\documentclass{elsart}
\usepackage{natbib}
\journal{SHPMP}

\begin{document}
\begin{frontmatter}
\title{Probability in modal interpretations of quantum mechanics}

\author{Dennis Dieks}
\address{Institute for the History and Foundations of Science\\
Utrecht University, P.O.Box 80.000 \\ 3508 TA Utrecht, The
Netherlands}
\ead{dieks@phys.uu.nl}
\date{January 2007}

\begin{abstract}
Modal interpretations have the ambition to construe quantum
mechanics as an objective, man-independent description of physical
reality. Their second leading idea is probabilism: quantum
mechanics does not completely fix physical reality but yields
probabilities. In working out these ideas an important motif is to
stay close to the standard formalism of quantum mechanics and to
refrain from introducing new structure by hand. In this paper we
explain how this programme can be made concrete. In particular, we
show that the Born probability rule, and sets of definite-valued
observables to which the Born probabilities pertain, can be
uniquely defined from the quantum state and Hilbert space
structure. We discuss the status of probability in modal
interpretations, and to this end we make a comparison with
many-worlds alternatives. An overall point that we stress is that
the modal ideas define a general framework and research programme
rather than one definite and finished interpretation.
\end{abstract}

\end{frontmatter}

\newpage

\section{Introduction: the modal point of view}
Modal interpretations of quantum mechanics are inspired by two
main ideas. The first  is to adopt a realist stance, in the
specific sense of interpreting the theory's mathematical formalism
in terms of properties and relations of physical systems,
independent of whether or not human observers are around. The
second motivating idea is that the relation between the formalism
of quantum theory and physical reality is to be taken as
\textit{probabilistic}. That is, according to modal
interpretations the quantum formalism does not tell us what
actually is the case in the physical world, but rather provides us
with a list of possibilities and their probabilities. The modal
viewpoint is therefore that quantum theory is about what
\textit{may} be the case---in philosophical jargon, quantum theory
is about \textit{modalities}.

This viewpoint is at odds with the operational viewpoint that the
quantum formalism should be interpreted  as a recipe for
predicting measurement outcomes. Modal interpretations strive for
a description of the world in terms of objective, man-independent
features, both on the macroscopic and (sub)microscopic level.
These features may turn out to be monadic properties of physical
systems (perhaps very exotic ones from a classical point of view),
or perhaps rather a structural network of relations or
perspectival properties; these are things to be decided during the
detailed elaboration of the interpretation.

In accordance with this modal philosophy, measurement results are
nothing but a subclass of the physical things existing in our
world: positions of pointers attached to measuring devices, marks
on computer tapes, etc. A measurement is a physical interaction
between an object system and a measuring device, and should be
treated like all other interactions. Modal interpretations
therefore only need the quantum formalism with unitary evolution,
i.e.\ the standard formalism without collapses. Indeed, as soon as
the idea that measurements are something special is abandoned, the
motivation for associating them with an own evolution mechanism,
collapses, disappears; only the unitary (Schr\"{o}dinger-like)
evolution remains. Modal interpretations thus fall into the class
of no-collapse interpretations of quantum mechanics.

So we assume that quantum mechanical states provide a description
of physical systems. It is good to be explicit here about the
distinction between the state as it is defined within the
mathematical formalism (a vector in Hilbert space, or a density
operator) and \textit{physical features of the represented
systems}---the latter are not mathematical entities. This
distinction is important because it is not an a priori obvious
matter what the exact relation between the mathematical state and
physical reality is; it is not self-evident in what way the
mathematical formalism represents. Discussions about the
interpretation of quantum mechanics sometimes neglect this point
by not clearly distinguishing between the formalism and what is
represented by it (e.g., by accepting as self-evident that a
$+$-sign in a superposition means \textit{joint existence}).

Our task in modal interpretations is thus to endow the standard
formalism, without collapses, with physical meaning. We need
interpretational rules that tell us how the mathematical formalism
relates to physical reality. Such rules do not constitute an
addition to the formalism of quantum mechanics: they are not part
of the mathematical formalism at all but establish a relation
between this formalism and the world. Any interpretation of
quantum mechanics will need to specify such a correspondence with
reality. Mathematical theories cannot fix their own
interpretation---as pieces of pure mathematics they do not contain
information about their possible applications.

Quantum mechanics has a familiar history of enormously successful
physical applications that make use of certain basic
interpretational rules that have proved their mettle; there is no
reason to doubt these. One of these basic interpretational rules
is that physical quantities are represented by hermitian operators
(observables). We will accept this standard correspondence (but
will change some other rules).

A natural form our interpretational question now takes is: which
physical quantities---represented by hermitian operators---can be
assigned a \emph{definite value}, when it is given that the
physical system is represented by a specific mathematical state.
Such definite values correspond to properties possessed by the
system. It might turn out in later developments that it is more
appropriate to focus on relations or perspectival properties
instead of the monadic properties represented by definite values
of physical quantities. But let us here focus on the standard
modal line, which relies on the attribution of properties in the
sense of definite values of physical quantities.

There may seem to be an easy answer to the question about the
relation between states and properties. Standard quantum mechanics
tells us that the state of a system is given by a density operator
$W$, obtained by `partial tracing' from the generally
\textit{entangled} state of the system and its environment. Now
consider $W$'s diagonal decomposition in terms of orthogonal
projections:
\begin{eqnarray*}
W & = & \sum_{i} p_{i}|\psi_{i}\rangle\langle\psi_{i}|,\\
& & \langle\psi_{i}|\psi_{j}\rangle = \delta_{ij}.
\end{eqnarray*}
\noindent This decomposition is unique if the coefficients $p_i$
are all unequal  (the case of non-uniqueness will be discussed
later on). There is a well-known way of interpreting such `mixed
states', namely via \textit{ignorance}: according to it the
physical system possesses one of the properties corresponding to
the projectors $|\psi_{i}\rangle\langle\psi_{i}|$, but we don't
know which one. In the special case of a pure state this reduces
to the standardly accepted eigenstate-eigenvalue rule, according
to which properties are only definite if the state is an
eigenvector of the corresponding projection operator.

However, there are well-known objections to the general validity
of this interpretation of $W$. The most important problem is that
if it were true that the partial system possessed the property
corresponding to $|\psi_{j}\rangle\langle\psi_{j}|$, then
according to the eigenstate-eigenvalue link the system should be
in the associated eigenstate $|\psi_{j}\rangle$. Analogously, the
other partial system (the environment) must be in a pure state as
well. But then a well-known theorem says that the total state must
be the product state of these two pure states. This is in conflict
with our initial assumptions: if the total state is
$|\psi_{j}\rangle \otimes |\xi_{j}\rangle$, or a mixture of such
states, there can be no entanglement whereas we assumed that in
general the total state \textit{is} entangled. The attribution of
one of the properties $|\psi_{j}\rangle\langle\psi_{j}|$ therefore
leads to contradictions.

To side-step this objection, modal interpretations propose to
\textit{drop the rule that a system can only possess a
well-defined value of a physical magnitude if it is represented by
an eigenstate of the corresponding observable}. In its stead comes
a new interpretative principle according to which the mathematical
state represents situations with definite physical properties even
if this state is a superposition of eigenstates of the
corresponding observables. The basic idea of interpreting the
formalism in this vein has been put forward, with a number of
variations, by several authors \citep{fra1,fra2,koc,die1,die2}.
Bas \citet{fra1}, who seems to have been the first to think along
these lines, coined the term `modal interpretation'; but we still
have to explain what the typically \textit{modal} aspects are. Let
us mention some more details in order to do so.

Consider the quantum mechanical treatment of a composite physical
system, consisting of two parts. In this case, the total Hilbert
space can be decomposed: $\mathcal{H} = \mathcal{H}_1 \otimes
\mathcal{H}_2$. According to a famous theorem (Schmidt,
Schr\"{o}\-din\-ger) there is a corresponding \textit{biorthogonal
decomposition} of every pure state in $\mathcal{H}$:
\begin{equation}
|\psi\rangle = \sum_{k} c_{k}|\psi_{k}\rangle\otimes
|R_{k}\rangle, \label{eq:modal} \end{equation} with
$|\psi_{k}\rangle$ in $\mathcal{H}_1, |R_{k}\rangle$ in
$\mathcal{H}_2$, $\langle\psi_{i}|\psi_{j}\rangle = \delta_{ij}$
and $\langle R_{i}|R_{j}\rangle = \delta_{ij}$. This decomposition
is unique if there is no degeneracy among the values of
$|c_{k}|^{2}$.

One well-known version of the modal interpretation gives the
following physical interpretation to this mathematical state. The
system represented by vectors in $\mathcal{H}_1$ possesses exactly
{\emph one} of the physical properties associated with the set of
projectors $\{|\psi_{k}\rangle\langle\psi_{k}|\}$, and definitely
does not possess the others. That is, exactly one of the mentioned
projectors is assigned the definite value $1$, the others get the
definite value $0$. The interpretation thus selects, on the basis
of the form of the state $|\psi\rangle$, the projectors
$|\psi_{k}\rangle\langle\psi_{k}|$ as definite-valued magnitudes.
All physical magnitudes represented by maximal hermitian operators
with spectral resolution given by $\Sigma a_{k}
|\psi_{k}\rangle\langle\psi_{k}|$ are also definite-valued, since
they are functions of the definite-valued projectors; their
possible values are given by the functions in question applied to
the values assumed by the projections.

In the case of degeneracy, that is $|c_{j}|^{2} = |c_{i}|^{2}$,
for $i,j \in I_{l}$ (with $I_{l}$ a set of indices), the
biorthogonal decomposition (\ref{eq:modal}) still determines a
unique set of projection operators, but these will generally be
multi-dimensional. The one-dimensional projectors have in this
case to be replaced by projectors $P_{l} = \sum_{i\in I_{l}}
|\psi_{i}\rangle\langle\psi_{i}|$. The physical properties now
correspond to this more general set of projectors. The general
class of definite physical quantities contains in this case
non-maximal hermitian operators in whose spectral resolution such
multi-dimensional projectors occur.

We have just stipulated that only \emph{one} of the values that
can be assumed by the definite-valued observables is actually
realized. This raises the question of what the \emph{probability}
is of the $l$-$th$ alternative being actual. In accordance with
the standard Born rule, this probability may be taken as
$|c_{l}|^{2}$ (in the case of degeneracy this becomes $\sum_{i\in
I_{l}} |c_{i}|^2$).

These details are mentioned here to clarify the spirit of the
modal ideas. We will devote a more fundamental discussion to their
justification, and to possible alternatives, later on. For
example, how does the notion that only \textit{one} possibility is
realized compare to the many-worlds alternative, according to
which \emph{all} terms in the superposition correspond to
actualities? And is it possible to derive and justify the Born
probability rule, instead of just positing it?

Taking for granted the probabilistic character of the
interpretation and the Born formula for the moment, we face the
consequence that, in general, the physical situation could have
been different from what it actually is, given the mathematical
state. Here we have the `modal' aspect of the interpretation: the
mathematical state does not fix what is actual, but specifies what
\emph{may} be the case. It follows from the probabilistic nature
of the relation between the state and the world that the same
physical situation may be realized `contingently' (if the
associated probability is smaller than one; things could have been
different in this case) or `necessarily' (if the probability is
one).

The just-explained interpretational rule ascribes definite
physical properties to physical sys\-tems, even if the state is a
superposition of eigenstates of the corresponding observables.
This has the following consequence. According to the von Neumann
measurement scheme, the situation after a measurement will
typically be described by a superposition of the form
(\ref{eq:modal}), with $|\psi_{k}\rangle$ denoting states of the
object system and $|R_{k}\rangle$ states of the measuring device
(`pointer position states'). The modal interpretation of this
state is that exactly one of the pointer positions is realized,
with a probability given by the Born rule. This is the modal
solution of the measurement problem: definite measurement outcomes
are predicted even though there are no collapses of the wave
function.

\section{Definite-valued observables}

Let us, after this review of the main ideas take a step back and
look in a more systematic way at the interpretational
possibilities that are left open by the motivating ideas behind
the modal interpretation. This will make it clearer to what extent
the just-discussed standard version can be justified, and will
also enable us to say something about the modal interpretation's
position among its competitors.

A core desideratum in devising the modal interpretation was the
wish to deny a special status to measurements: measurements should
be dealt with in the same way as ordinary physical interactions.
Combined with the desire to keep the usual quantum formalism
intact, this leads to a specific class of interpretations, namely
no-collapse interpretations, in which there is only unitary
evolution in Hilbert space. The task of all these interpretations
is to link the unitarily evolving states in Hilbert space to
physical features of the represented systems. The simplest
programme for accomplishing this consists in the attempt to define
a set of definite-valued observables from the mathematical state.
These are observables that can be assigned a well-defined
numerical value, and that thus fix physical quantities. Now, it is
a notorious feature of the Hilbert space formalism that not all
observables (i.e., hermitian operators) can be assigned definite
values simultaneously (if we respect the functional relations
between them); this follows from the Kochen-Specker theorem. The
question therefore arises: What is the maximum set of observables
that are definable from the quantum state and can jointly be given
definite values without getting into contradictions? The usual
quantum---Born---probabilities should become expressible as
probability distributions over these definite values. Indeed, as
already emphasized, measurement results are special cases of
system properties in this approach, and the Born probabilities
therefore come to pertain to the values of physical
quantities---in the same way as probabilities in classical phase
space.

In our search for definite-valued observables it is possible to
include interpretations like the Bohm interpretation if we allow
for the possibility that there is a preferred observable $R$ that
is always definite, for all quantum states (in the Bohm theory
position plays this role). The situation in which no privileged
observable exists then becomes a special case. It should be noted,
however, that there is a tension between this assumption of a
preferred observable and the desideratum that the ordinary quantum
formalism should be retained as much as possible: in the usual
Hilbert space formalism there is no preferred observable. After
discussing the case with a preferred observable we will therefore
focus on the alternative, in which there is no such \emph{a
priori} fixed physical quantity, and in which the definite-valued
observables are determined by the quantum state alone.

Consider an arbitrary pure quantum state represented by a ray
$\psi$ in the Hilbert space $\mathcal{H}$. Let the Boolean algebra
generated by the eigenspaces of the preferred observable $R$ be
denoted by $\mathcal{B}(R)$. The usual quantum mechanical
probabilities of the values of $R$, calculated via the Born rule
applied to $\psi$, can be represented by an ordinary Kolmogorovian
measure over the 2-valued homomorphisms (consistent assignments of
truth values $0$ and $1$) on $\mathcal{B}(R)$. We now ask for the
maximal lattice extension $\mathcal{D}(\psi, R)$ of
$\mathcal{B}(R)$, formed by adding eigenspaces of other
observables, such that we can represent in the same way the Born
probabilities both for values of $R$ and for the values of these
other observables.

Since we do not want to accept more mathematical structure not
automatically present in Hilbert space than already introduced by
the presence of $R$ as a preferred observable, we require these
definite-valued observables to be definable solely in terms of
$\psi$ and $R$. It follows that each element of $\mathcal{D}(\psi,
R)$ should be invariant under all automorphisms of Hilbert space
that preserve both the ray $\psi$ and the eigenspaces of $R$. This
requirement \citep{dieks5} is slightly stronger than the one made
in \citep{bub0,bub,bub2}; we shall comment on the difference
below. This invariance, expressing definability from $\psi$ and
$R$, will do most of the work in determining our definite-valued
observables.

Let us consider an $n$-dimensional Hilbert space $\mathcal{H}$,
and an observable $R$ with $m \leq n$ distinct eigenspaces $r_{i}$
of $\mathcal{H}$. Let $\psi_{r_{i}}$, $i = 1,2, \ldots, m$, denote
the orthogonal projections of $\psi$ onto these eigenspaces
$r_{i}$. Now, the set of automorphisms that leave $\psi$ and $R$
invariant includes all automorphisms that are equal to the unity
operator when they operate on vectors orthogonal to $r_{i}$, and
are rotations around $\psi_{r_{i}}$ or reflections with respect to
the space orthogonal to $\psi_{r_{i}}$ inside $r_{i}$---since
these transformations leave the eigenspaces of $R$ and the
projections of $\psi$ the same, $R$ and $\psi$ themselves are
invariant. Now consider a projection operator $P$ that is to
correspond to a definite-valued property; so $P$ must be definable
from $\psi$ and $R$. If the subspace of $\mathcal{H}$ on which $P$
projects is contained in one of the $r_{i}$, this subspace should
therefore be invariant under the mentioned rotations and
reflections with respect to $\psi_{r_{i}}$. This leaves four
possibilities for the subspace in question: it can be the
null-space, $\psi_{r_{i}}$, $\psi_{r_{i}}^{\perp} \wedge r_{i}$,
or $r_{i}$.

In the general case, the subspace on which $P$ projects will not
be contained in one of the $r_{i}$ spaces, but will have non-zero
projections on a number of them. The requirement that the subspace
remains invariant under the above-mentioned automorphisms now
implies that its projections on the different spaces $r_{i}$ are
each either null, $\psi_{r_{i}}$, $\psi_{r_{i}}^{\perp} \wedge
r_{i}$, or $r_{i}$. All possible subspaces on which $P$ may
project are therefore found by taking one of these latter spaces
for each value of $i$, and constructing their span.

The lattice of subspaces that may correspond to definite
propositions is therefore generated by all sublattices
$\{0,\psi_{r_{i}}, \psi_{r_{i}}^{\perp} \wedge r_{i},
r_{i}\}$\footnote{The set of automorphisms that leave $\psi$ and
$R$ invariant contains more elements than the ones considered in
this derivation---so what has been proved is that the lattice of
definite properties cannot be larger than the one constructed
here. As the constructed lattice is clearly definable from $\psi$
and $R$ and satisfies the other requirements, it is the maximal
lattice we were looking for.}. In the case that $r_{i}$ is
one-dimensional, $\psi_{r_{i}}$ is equal to $r_{i}$ and
$\psi_{r_{i}}^{\perp} \wedge r_{i}$ equals $0$, so that the
sublattice reduces to $\{0,r_{i}\}$.

It is clear from this construction that the resulting set of
definite-valued projection operators is indeed a lattice: it is
closed under the lattice operations of disjunction and conjunction
(corresponding to taking the span or intersection of the
associated eigenspaces). Moreover, the lattice is Boolean: all
projection operators in it commute with each other. Therefore, no
Kochen and Specker-type paradoxes can arise, and the quantum
mechanical probabilities (including joint probabilities) can be
represented by means of a classical measure on the lattice.

The above construction made use of the existence of a preferred
observable, namely $R$. As stated above, it is important to see
what happens if the state $|\psi \rangle$ and the Hilbert space
structure are the only entities used to define the definite
properties.

A possible way of implementing this is to take the projection on
$|\psi \rangle$ itself for $R$. If we denote the subspace
orthogonal to $\psi$ by $\psi^{\perp}$, we obtain the definite
lattice consisting of the subspaces $\{0, \psi, \psi^{\perp},
\mathcal{H}\}$. The same result is obtained if we take the unity
operator on $\mathcal{H}$ for $R$. This therefore leads to the
`orthodox' property assignment: only observables of which
$|\psi\rangle$ is an eigenvector qualify as definite-valued
\citep{bub}. This traditional way of assigning properties returns
us to the measurement problem, because after a measurement the
combined system of measuring device and object system ends up in
an entangled state that according to this assignment does not
correspond to a definite pointer property. However, on second
thought the situation is more complicated. The projection operator
$|\psi\rangle \langle\psi|$ is an observable of the \textit{total}
system, and the just-mentioned property assignment pertains
likewise to this total system. But we are really interested in the
\textit{individual} properties of device and object taken by
themselves. Therefore, we need substitutes for
$|\psi\rangle\langle\psi|$ that represent the states of these
individual systems. In the context of standard quantum mechanics
such operators are readily available, namely the density operators
for the partial systems. Postponing for a moment possible doubts
about the status of these operators in this new context, we are
thus prompted to consider the definite lattices that result if the
operators $W_{1}\otimes I$ and $I\otimes W_{2}$ are taken for $R$
(the total Hilbert space is the tensor product of Hilbert spaces
belonging to the partial systems, $\mathcal{H} = \mathcal{H}_{1}
\otimes \mathcal{H}_{2}$).

Denote the eigenspaces of $W_{1}\otimes I$ by $w_l \otimes
\mathcal{H}_{2}$, $l= 1,2,\ldots$. Now write $|\psi\rangle$ as a
biorthogonal decomposition
\begin{equation}
|\psi\rangle = \sum_{k,j} c_{k,j}|\alpha_{k,j}\rangle\otimes
|\beta_{k,j}\rangle, \label{eq:modal1} \end{equation} with
$|\alpha_{k,j}\rangle$ in $\mathcal{H}_1, |\beta_{k,j}\rangle$ in
$\mathcal{H}_2$, $\langle\alpha_{l,i}|\alpha_{m,j}\rangle =
\delta_{lm}.\delta_{ij} = \langle \beta_{l,i}|\beta_{m,j}\rangle$.
The second index, $j$, takes possible degeneracies into account:
$|c_{k,j}|^{2}$ depends only on $k$, not on $j$. The projection of
$|\psi\rangle$ on $w_l \otimes \mathcal{H}_2$ is given by
$|\psi_{l}\rangle= \sum_{j} c_{l,j}|\alpha_{l,j}\rangle\otimes
|\beta_{l,j}\rangle$. As we have seen, it follows that the lattice
of definite properties is generated by the sublattices $\{0,
\psi_{l}, \psi_{l}^{\perp} \wedge (w_l \otimes \mathcal{H}_2),
w_{l} \otimes \mathcal{H}_2\}$. We can restrict this lattice to a
lattice of definite properties of the first system alone
(represented in $\mathcal{H}_1$) by looking for those definite
projections in the lattice that possess the form $P \otimes I$.
The projection operators $P$ can in this case be taken to
represent properties of system $1$ by itself. Inspection of the
lattice shows that all projections of the sought form are
generated by the projectors $P_{w_{l}} \otimes I$. The restriction
of the lattice of definite properties of the combined system to a
lattice of definite properties of system $1$ is therefore the
Boolean lattice generated by the projections $P_{w_{l}}$. These
are exactly the properties assigned by modal interpretations of
the type discussed in the Introduction \citep{die1,die2,vermaas2}.
In measurement situations these definite properties should
correspond to pointer positions.\footnote{It is an important
question whether this requirement of empirical adequacy is in fact
fulfilled. In situations with a limited number of degrees of
freedom this has been shown to be the case \citep{bacciahemmo};
but there are grounds for doubt in cases in which the number of
degrees of freedom is infinite or very large \citep{baccia}; see
\citep{benedie} for a possible response).}

The analysis just given is similar to the one proposed by Bub and
Clifton \citep{bub0,bub,bub2}. The difference is that these
authors required the \textit{set} of definite properties
\textit{as a whole} to be definable from $|\psi\rangle$ and $R$,
whereas here we have imposed the stronger demand that the
\textit{individual definite properties} be so definable. Given the
idea that $|\psi\rangle$ should fix as many elements of the
interpretation as possible, our stronger requirement seems the
more natural one; moreover, it makes the analysis considerably
simpler. As was to be expected, the lattice of definite properties
that we found above on the basis of our stronger requirement is
included in the lattice determined by Bub and Clifton. The latter
possesses more `fine structure': The Bub-Clifton lattice contains
projection operators that cannot be defined individually but still
belong to the set of projectors defined as a whole. However, these
differences are not very significant. In the case of the
measurement-like situation we have just discussed, the only
difference is that in the Bub-Clifton approach all individual
one-dimensional projections within the null-space of $W_{1}$ are
definite, whereas in our approach it is only the total projector
on this null-space that is definite-valued.

In this derivation we took $W_{1}$ for the state of system $1$.
This is standard practice in quantum mechanics; however, the usual
justification relies on the probabilistic interpretation of the
theory and the Born rule. It would be preferable not to presuppose
anything about this at the present stage. Indeed, in the next
section we will make an attempt to derive the Born rule. It is
therefore desirable to have a derivation of the definite-valued
observables of the partial systems that does not presuppose that
the density operators $W_{i}$ characterize the individual systems.

In order to achieve this we again make use of the biorthogonal way
of writing $|\psi \rangle$, Eq.\ (\ref{eq:modal1}). As before, our
aim is to determine maximal sets of properties of system $1$ that
can be defined from this state. We will use that $\mathcal{H}$ is
the tensor product of the Hilbert spaces of the individual systems
$1$ and $2$, $\mathcal{H} = \mathcal{H}_1 \otimes \mathcal{H}_2$:
we want the definite properties of system $1$ to be invariant
under automorphisms that leave $|\psi \rangle$ the same and that
respect this factorization of the total Hilbert space. These
automorphisms have the form $U_{1} \otimes U_{2}$, with $U_{1}$
and $U_{2}$ defined on $\mathcal{H}_1$ and $\mathcal{H}_2$,
respectively \citep{dieks4,dieks5}. So we ask which automorphisms
$U_{1} \otimes U_{2}$ leave $|\psi \rangle$ invariant, and which
projectors in $\mathcal{H}_1$ remain the same under their
operation (in other words, under the operation of the associated
$U_{1}$).

To investigate this we must have a closer look at the invariance
properties of (\ref{eq:modal1}). This biorthogonal decomposition
is unique up to certain unitary transformations \textit{within}
the subspaces spanned by the vectors $\{|\alpha_{k,j}\rangle\}_j$
(and $\{|\beta_{k,j}\rangle\}_j$), with fixed values of $k$ (these
are the `degeneracy subspaces', labelled by values of $k$). This
can be seen in the following way. The component of $|\psi \rangle$
\textit{within} such a degeneracy subspace can (after
normalization) be written as
\begin{equation}
|\omega\rangle = \sum_{j} N^{-1/2}
\exp{i\phi_j}|\alpha_{j}\rangle\otimes |\beta_{j}\rangle,
\label{eq:modal2} \end{equation} with $N$ the dimension of the
subspace in question. Now take an arbitrary unitary operator $U_I$
in the subspace of $\mathcal{H}_1$ spanned by the vectors
$|\alpha_{j}\rangle$. Define an operator $U_{II}$ in the subspace
of $\mathcal{H}_2$ spanned by the vectors $|\beta_{j}\rangle$,
through its matrix elements, as follows:
\begin{equation}
\langle \beta_k|U_{II}|\beta_l\rangle = \overline{\langle \alpha_k|U_{I}|\alpha_l\rangle}. \exp{i(\phi_k - \phi_l)},
\label{eq:u2}
\end{equation}
with the bar denoting complex conjugation. It follows from this
definition that $U_{II}$ is unitary (given the unitarity of
$U_I$). We can now construct a product unitary operator in the
tensor product of the two subspaces: $U= U_I \otimes U_{II}$. This
operator leaves $| \omega\rangle$ invariant:
\begin{eqnarray}
\langle \omega|U|\omega\rangle = N^{-1}\sum_{i,j}\exp{i(\phi_j - \phi_i)}\langle \alpha_i|U_{I}|\alpha_j\rangle
\langle \beta_i|U_{II}|\beta_j\rangle = \nonumber \\ =  N^{-1}\sum_{i,j}|\langle \alpha_i|U_{I}|\alpha_j\rangle|^2 = 1 .
\end{eqnarray}
In other words, we can operate with an arbitrary unitary operator
$U_I$ in one of the degeneracy subspaces of $\mathcal{H}_1$, and
undo its effect on $|\psi\rangle$ by operating with a suitably
chosen unitary operator $U_{II}$, as defined in Eq.\
(\ref{eq:u2}), in the corresponding degeneracy space of
$\mathcal{H}_2$. It is of course also true that operating in a
similar way with an arbitrary unitary in a degeneracy subspace of
$\mathcal{H}_2$ can be undone by a corresponding unitary operation
in $\mathcal{H}_1$.

In the case of a one-dimensional subspace (i.e., corresponding to
a non-degenerated term in the superposition), $U_I$ can only be a
multiplication by a phase factor, $\exp{i\phi}$. In this case the compensating
$U_{II}$ takes the form of multiplication by the inverse factor,
$\exp{-i\phi}$.

Any spaces contained within the degeneracy subspaces selected by
the biorthogonal decomposition are clearly not invariant under all
unitary product operations in $\mathcal{H} = \mathcal{H}_1 \otimes
\mathcal{H}_2$ that preserve $|\psi \rangle$; only the degeneracy
subspaces themselves are invariant. These spaces are exactly the
eigenspaces of the reduced density operator $W_{1}$.

So we arrive in a quick and simple way at the same conclusion as
before: the lattice of those properties of system $1$ that can be
defined on the basis of $|\psi \rangle $ alone, is generated by
the projection operators $P_{w_{k}}$. Since this lattice is
Boolean, definite values can be jointly assigned to all its
elements without contradictions, and measures on the lattice can
be represented in a classical Kolmogorovian probability space.

We have thus found a uniqueness result for the set of
definite-valued observables. But any uniqueness result stands or
falls with its premises. In our derivation we assumed that the
definite-valued observables should be definable from $|\psi
\rangle$ and the splitting of the total Hilbert space into two
factor spaces, representing the system and its environment,
respectively. If it is assumed that more or other ingredients play
a role in determining the properties of a system, different
definite observables will result. It has been suggested, for
example, that we should not just look at the system and its
environment, but rather at a three-fold splitting of the total
Hilbert space into factors corresponding to the system, a
measuring device (or, more generally, a system that is able to
make records) and the remaining environment, respectively
\citep{zur1,schlosshauer}. If the total state can be written as
\begin{equation}
|\Psi_{sA\varepsilon} \rangle = \sum_k a_k |s_k \rangle |A_k \rangle
|\varepsilon_k\rangle, \label{3split}
\end{equation}
with $s$, $A$ and $\varepsilon$ referring to system, device and
environment, respectively, with orthogonal pointer states $\{|A_k
\rangle \}$, these orthogonal pointer states will be the only ones
that make the three-fold factorization of (\ref{3split}) possible.
In this way a set of preferred pointer states of the device can be
defined. A problematic feature of this proposal is that it does
not lead to a general assignment of properties to arbitrary
systems; and that states of the form (\ref{3split}) are very
special. Such states will only result from specific interactions
\citep[sect.\ 4.2]{zur1}.  In the context of decoherence studies
it has indeed often been suggested that it is the form of the
\textit{interaction Hamiltonian} between (macroscopic) systems and
their environments that does the selecting of preferred states:
that pointer states are `memory states' that behave in a robust
and approximately classical way. The selected pointer observable
commutes with the interaction Hamiltonian (perhaps in an
approximate way), so that the environment effectively performs a
non-demolition measurement of the pointer observable---see also
\cite{zur}.\footnote{\citet[sect.\ IIB]{zur1} states, however,
that all `measurable properties' of a system can depend only on
its own state, obtained by partial tracing from the entangled
state of the system and its environment. This seems to lead us
back to the standard modal property attribution.} These proposals
deserve further study. Questions to be asked are, e.g., about the
presence of physical properties in cases without suitably
interacting environments and recording devices. Can sense be made
of the suggestion that the concept of physical properties becomes
only applicable in special circumstances? Another issue is the
status of the approximations that are usually involved in
decoherence calculations (compare section 5 below). Anyway, it has
to be admitted that decoherence proposals have led to plausible
candidates for preferred states in many model calculations.

It should therefore be stressed that the modal ideas constitute a
research programme rather than a completely fixed interpretation.
The central features remain that the quantum formalism describes
the world in man-independent terms, in particular without
according a special role to measurements undertaken by humans; and
that the relation between formalism and physical reality is
probabilistic. This leaves room for differences in detailed
elaborations.

\section{The Born measure}

The modal interpretation is probabilistic and must therefore
define a probability measure on the lattice of definite-valued
observables. This raises the question: Is it possible to derive a
preferred measure on the lattice of definite-valued observables,
along the same lines as in the derivation of the definite-valued
observables? More specifically: if we impose the requirement that
the measure is to depend only on the state in Hilbert space, the
tensor product structure of Hilbert space and the preferred
observables induced by the state, has this enough bite to single
out a definite form of the measure? An affirmative answer would
fit in nicely with the modal philosophy according to which the
standard quantum mechanical formalism is descriptively complete;
that no elements need to be added by hand.

As we will argue, the answer is `yes': the Born measure is the
only one that is definable from just the relation between $|\psi
\rangle $ and its associated definite-valued observables.

Denote the measure to be assigned to the definite-valued projector $P$,
if the state is $|\psi \rangle $, by $\mu(|\psi\rangle, P)$. Write
$|\psi\rangle$ in its biorthogonal form again:
\begin{equation} |\psi\rangle = \sum_{k}
c_{k}|\alpha_{k}\rangle\otimes |\beta_{k}\rangle,
\end{equation} where we now have taken the non-degenerate case for
simplicity. First note that we can take the coefficients $c_{k}$
to be real numbers: all phase factors can be absorbed into the
vectors $|\alpha_{k}\rangle$ or $|\beta_{k}\rangle$, without any
effect on the observables that are value-definite (the projection
operators are invariant under this operation). So if $\mu$ is
going to depend on the coefficients $c_{k}$, only their absolute
values or, what amounts to the same thing, only $|c_{k}|^2$ can
enter the expression.

An alternative road to this conclusion is to use the
transformations $U_I \otimes U_{II}$ under which $|\psi \rangle$
is invariant. As we have seen in the previous section, both $U_I$
and $U_{II}$ are pure phase transformations in this non-degenerate
case. One could now reason as follows, like \citet{zur1}: any
physical features pertaining to system $I$ alone should be
invariant under the operation of any $U_I$ on $|\psi \rangle$, for
the following reason. The effect of $U_I$ can be undone by
$U_{II}$ ($U_I$ is what Zurek calls an `envariance' operation);
and $U_{II}$ should not be expected to affect the physical
properties of $I$. Consequently, any effects $U_I$ may have on the
mathematical state of $I$ should not be relevant to the physical
features of $I$. In particular, the phases of the coefficients
$\{c_k\}$ must be irrelevant, so that only their absolute values
can count.

To conclude that $\mu$ indeed only depends on $\{c_{k}\}$, we need
an additional argument, however. As pointed out by \citet{caves},
it is not `envariance' that is doing the work here: rather, the
assumption (also made by Zurek) that the probabilities and
physical properties pertaining to system $I$ do not depend on the
vectors $\{|\beta_{k}\rangle \}$ in the biorthogonal decomposition
is central---and once we make this assumption, the notion of
envariance is no longer needed. This assumption may be seen as a
no-signalling condition: its violation would make it possible to
change physical features of $I$ by intervening in the state of
$II$, which would make it possible to signal. It can also be
regarded as a non-contextuality condition: it should not make a
difference for the characteristics of $I$ what unitary operations
are taking place in its environment $II$. This entails invariance
of the probabilities under arbitrary $U_{II}$. In particular,
$\mu$ can only depend on system $I$'s definite-valued projectors
$P$ and on $\{c_k\}$, and since absorbing all phase factors into
$\{|\beta_{k}\rangle\}$ does not change the probabilities only the
absolute values of $\{c_k\}$ can be relevant---for further
discussions of Zurek's line of argument see
\citep{barnum,caves,mohrhoff,schlosshauer1}. In brief, the
non-contextuality condition entails that all probabilities must be
invariant under application of arbitrary unitaries $U_I \otimes
U_{II}$. So we find that the bases $\{|\alpha_{k}\rangle\}$ and
$\{|\beta_{k}\rangle\}$ are irrelevant for the probabilities, and
only the values $\{|c_k|\}$ can play a role.

However, this irrelevance of $\{|\alpha_{k}\rangle\}$ and
$\{|\beta_{k}\rangle\}$ for the expression of $\mu$ can be
justified in a more direct way, without invoking principles about
causality and contextuality, by a definability argument like the
one in the previous section. We want $\mu$ to be definable
exclusively from $|\psi\rangle$ and the product Hilbert space
structure. We can therefore immediately impose the requirement
that unitary transformations of the form $U_I \otimes U_{II}$
should not change the values taken by the measure; that these
values remain the same, but now apply to transformed projectors
(like ${U_I}^{-1}PU_I$). The reason is that these unitary
transformations only change the orientation of $|\psi\rangle$ in
Hilbert space, but do not change anything in the relation between
$|\psi\rangle$ and the definite-valued observables determined by
it; all changes are equivalent to those induced by a basis
transformation in the Hilbert space, and can be undone by
performing an inverse basis transformation. But we want $\mu$ to
be determined solely by the the state and its associated
definite-valued projectors---the choice of a basis in Hilbert
space in terms of which the state is written down should be
immaterial. In other words, the same collection of $\mu$ values
must be associated with the entire class of states that follow
from $|\psi\rangle$ by application of arbitrary unitary operations
$U_I \otimes U_{II}$. Since the only feature that is common to all
these states are the values of $|c_i|$, $\mu$ must be a function
of these values only. As pointed out above, we can therefore
consider $\mu$ to be a function of $\{|c_i|^2\}$.

Now compare the situation described by $|\psi\rangle$ with the one
in which we discard, forget, or are unable to observe the
differences between the different $|\beta_{k}\rangle$ for
$k\geq2$. The total probability of not having $|\beta_{1}\rangle$
should now be the sum of the probabilities of $|\beta_{k}\rangle$
for $k\geq 2$, since distinct alternatives have been grouped
together. The vector that would correspond exactly to this new
situation results from $|\psi\rangle$ by erasing the differences
between $|\beta_{k}\rangle$ for $k\geq 2$, and replacing all these
vectors by $|\beta_{2}\rangle$. This leads to the state
\begin{equation} |\chi\rangle = c_{1}|\alpha_{1}\rangle\otimes
|\beta_{1}\rangle + \sqrt{\sum_{k=2}|c_k|^2}|\alpha\rangle\otimes
|\beta_{2}\rangle,
\end{equation} where $|\alpha\rangle$ is a normalized vector. The measure
assigned by this state to $|\beta_2\rangle \langle\beta_2 |$
should be the sum of the original measures of the projectors that
have coalesced into $|\beta_2\rangle \langle\beta_2 |$.

Finally, because $\sum_{k=2}|c_k|^2 = 1 - |c_1|^2$ we may write
$\mu(|\beta_1\rangle \langle\beta_1 |)= f(|c_1|^2)$. By parity of
reasoning we may write down an analogous formula for the other
projectors: $\mu(|\beta_i\rangle \langle\beta_i |)= f(|c_i|^2)$.

On the basis of our above observation about the relation between
the measures induced by $|\psi\rangle$ and $|\chi\rangle$,
respectively, we now find that \begin{equation} f(\sum|c_k|^2)=
\sum f(|c_k|^2).
\end{equation}
From this it follows that $ f(|c_k|^2) = const. |c_k|^2$, and in
view of normalization
\begin{equation} \mu(P_k) = |c_k|^2.
\end{equation} This is the Born rule.

\section{Probability and modality}

As explained in the Introduction, modal interpretations understand
$\mu$ as a probability: given the state $\psi$ in Hilbert space,
exactly \textit{one} of the projectors that are singled out as
definite-valued by $\psi$ possesses the value $1$, and the chance
that this value is taken by $P_k$ is given by $|c_k|^2$. In
general there are more than one possibilities for the actual
physical situation (defined by the values of the definite-valued
observables), once the state in Hilbert space has been given; the
state specifies a probability distribution over them. This
probability quantifies our ignorance about the actually obtaining
physical situation in cases in which we know the state in Hilbert
space and have no additional information. It is also reflected in
the relative frequencies with which physical properties occur in
repetitions of situations corresponding to the same $\psi$. In
other words, the probabilities occurring in the modal
interpretation have the same status as classical probabilities and
have the usual classical interpretations. That $\mu$ has this
physical meaning in terms of probabilities and ignorance is
clearly something that is not decided by the mathematical
formalism itself (see for a dissenting voice \citep{zur1}, and for
a convincing critical analysis of this argument e.g.\
\citep{mohrhoff}). It is an interpretational postulate that should
be judged on the basis of comparison with alternatives---we shall
have more to say about this in the next section.

According to the modal interpretation the state in Hilbert space
thus is about possibilities, about what may be the case; about
modalities. But there is also a second aspect to $\psi$: it is the
theoretical quantity that occurs in the evolution equation, and
its evolution governs deterministically how the set of definite
valued quantities changes. This double role of $\psi$, on the one
hand probabilistic and on the other dynamical and deterministic,
is a well-known feature of the Bohm interpretation. As we have
seen in section 2, the Bohm interpretation can be regarded as a
specific version of the modal interpretation, namely one in which
there is an \textit{a priori} given definite-valued observable. As
we see now, the double deterministic-and-probabilistic aspect of
$\psi$ is typical of modal interpretations quite generally.

\section{One versus many worlds}

The no-collapse scheme by itself does not imply anything about
probability: it just says that the Hilbert space state evolves
unitarily. An interpretation, which is external to the formalism,
must be supplied before anything can be stated about what the
state represents. It is sometimes suggested, however, in
opposition to this, that the no-collapse formalism is capable of
providing its own interpretation \citep[p.\ 168]{dewitt}. What
seems to be meant is the claim that there exists a
\textit{simplest} interpretation that does most justice to the
symmetries inherent in the Hilbert space formalism. In particular,
the suggestion is that, granted the usual interpretational links
between eigenstates of observables and values of physical
quantities, a superposition of such eigenstates should be
interpreted as representing the joint existence of the
corresponding values. This is the many-worlds idea: superpositions
represent collections of worlds, in each one of which exactly one
value---corresponding to one term from the superposition---of an
observable is realized. The claim is that this many-worlds
interpretation distinguishes itself by being simple and by
possessing a natural fit to the formalism, respecting its
symmetries.

In a superposition all terms occur in the same way, i.e.\ without
any markers that single out one, or some, terms as corresponding
to what actually is the case. The basic thought of the many-worlds
interpretation is that this symmetry signifies that all terms
correspond to reality in the same way: if one term refers to
something actually existing, then so must all. The identification
of any particular term as representing actuality is regarded as
breaking the symmetry present in the state, and therefore as
objectionable.

Let us have a closer look at this argument. It may be conceded
that singling out any particular term from a superposition, and
identifying it as the one referring to actuality, breaks the
symmetry of the state. But do probabilistic interpretations really
work this way; do they single out one term over the others?
Consider the analogous situation in classical probability theory:
the same train of thought applied there would also lead to the
conclusion that all events to which a probability distribution
assigns a value should be simultaneously realized, if we do not
have an underlying deterministic theory. One would be led, also in
the classical case, to a many worlds ontology as the one that best
fits the probability formalism. But is this interpretation really
simpler or more symmetric than the usual one? Answering this
question requires comparing two different mappings (reference
relations), with a mathematical event space as their common
domain. The probabilistic mapping is from the event space to
\textit{possibilities}; whereas the many-worlds mapping maps all
elements of the event space into \textit{realities}. Apart from
this difference in status of the elements of the ranges of the two
mappings (possibility and reality), which as far as the mapping
itself is concerned is just a difference in labels, everything is
the same. It is therefore hard to see how there could be any
difference in simplicity, naturalness or symmetry.

The impression that there nevertheless is such a difference
evidently derives from the notion that the probabilistic
interpretation identifies one of the possibilities as the actual
one, and thus violates the symmetry that is present in the
many-worlds option. But this notion is incorrect. \textit{Not}
singling out such a privileged event is precisely what makes an
interpretation fundamentally probabilistic. The probabilistic
option treats all elements of the probability space in exactly the
same way, by mapping them to possibilities that \textit{may} be
realized---it does not tell us which possibility \textit{is}
realized. Each single element of the interpretation's range may
correspond to reality. There is therefore the same symmetry as in
the many-worlds option.

Still, there is a difference. In the probabilistic interpretation
it is stipulated from the outset that exactly one possibility is
realized. Even if there is symmetry with respect to which
possibility this is, is it not true that this one-world
stipulation by itself introduces surplus structure that is not
present in the many-worlds interpretation? I do not think this is
right. There is perfect equivalence in the sense that the
many-worlds interpretation is defined by the condition that each
element of the measure space corresponds to an actual states of
affairs, whereas the probabilistic alternative is defined by the
condition that each element may correspond to the one actual (but
unspecified) state of affairs. There is consequently no difference
in the symmetry properties or simplicity of the interpretations,
but rather a difference in the nature of their ranges: in the one
case this is a collection of many real worlds, in the other it is
a collection of candidates for the one real world. So, in the end
the significant difference boils down to the difference between
one and many---and it surely is not a principle of metaphysics or
rational theory choice that many is simpler than one. General
considerations concerning symmetry and simplicity do therefore not
favor a many worlds interpretation over a probabilistic, modal,
interpretation.

Let us briefly discuss a further general problem with the
many-worlds idea, namely the well-known question of how to
accommodate the notion of probability at all in a theory according
to which it is certain that all possibilities will be actually
realized. The dominant opinion among many-worlds adherents seems
to have become that the quantum probabilities should be seen as
subjective, in the sense of quantifying a subject's degree of
belief about the future experiences of his splitting self (though
not subjective in the sense of purely personal: The Born
probabilities should come out as governing the objectively most
rational choices). This Deutsch-Wallace line of argument
\citep{deutsch, wallace,wallace1} proceeds from the assumption
that a subject should be indifferent between terms in the total
superposition that occur with equal `weights', i.e.\ squared
absolute values of the coefficients. This apparently
\textit{presupposes} a probabilistic conception of the quantum
state---even though it is now probabilistic in the subjective
sense. Indeed, it is \textit{a priori} unclear why there should be
unique rational expectations defined \textit{at all} in a
situation corresponding to a particular $\psi$ if we do not start
out by assuming a probabilistic meaning of the wave function. And
even if we do accept that there are measures of our credence
hidden in the quantum formalism, it still is not self-evident that
the symmetries of the quantum state are significant for them (the
latter point is also made by \citet{price}). In our modal approach
we did not face these problems, because we explicitly took an
interpretational step and \textit{postulated} a probabilistic
meaning of $\psi$, and moreover required the probabilities to be
definable in terms of $\psi$. It is this latter requirement that
makes the symmetries in $\psi$ relevant for the probability
assignment.

That the universe consists of many actual worlds, each one
containing exactly one possible outcome of a process, actually
does not play a role in the technical part of the Deutsch-Wallace
argument. It is only the quantum state $\psi$ that enters into the
reasoning, as said with the assumption that $\psi$ should govern
rational expectations to start with. Any conclusions that can be
drawn from such reasoning in the context of interpretations with
many coexisting actual worlds, can surely also be drawn in the
context of the more usual probabilistic construal (namely that
only one possibility will be actually realized), or so it would
seem. However, \citet{wallace1} argues that the many worlds
interpretation \textit{is} essential here. In the course of
defending the idea that a rational agent should be indifferent
between outcomes that occur with equal weights in the
superposition $\psi$ (he calls this principle
\textbf{equivalence}), Wallace says:
\begin{quote}
``I wish to argue that the Everett interpretation necessarily
plays a central role in any such defence: in other
interpretations, \textbf{equivalence} is not only unmotivated as a
rationality principle but is actually absurd.

Why? Observe what equivalence actually claims: that if we know
that two events have the same weight, then we must regard them as
equally likely regardless of any other information we may have
about them. Put another way, if we wish to determine which event
to bet on and we are told that they have the same weight, we will
be uninterested in any other information about them.

But in any interpretation which does not involve branching---that
is, in any non-Everettian interpretation---there is a further
piece of information that cannot but be relevant to our choice:
namely, \textit{which event is actually going to happen}? If in
fact we know that E rather than F will actually occur, \textit{of
course} we will bet on E, regardless of the relative weights of
the events''.
\end{quote}
It seems to me that this argument does not work. In reasoning
about rational expectations---connected to subjective
probabilities---about future events one standardly distinguishes
`admissible' from `inadmissible' information. For example, David
Lewis's `Principal Principle' says that a rational agent should
set his subjective probability equal to the objective chance of an
event, unless there is (inadmissible) information about what is
actually going to happen \citep{davidlewis}. In other words, it is
not a sound principle that if two events have the same objectively
and rationally founded subjective probability, then we have to
regard them as equally likely completely regardless of any other
information we may receive. In our probability judgements we do
not use information about the actual outcomes; if such information
were to reach us we would of course adapt our expectations in
spite of the probabilities. Conversely, no principle saying that
our expectations should not change whatever further information
reaches us should guide our search for the values of
probabilities. If it did, no probability values other than $1$ and
$0$ could ever be assigned, because in principle we might always
receive information (e.g., by revelation) about whether things
will or will not happen. This remains the case in the many-worlds
scenario. It is true that all possibilities will become actual in
this scenario, so that we cannot learn which event is actually
going to happen. But we may still receive information about our
\textit{actual future experiences}, which according to Wallace are
subject to uncertainty in the many-worlds universe because we
ourselves split and do not know who of our successors we will
become (this way of accommodating uncertainty and probability in
the many-worlds interpretation is itself the subject of
controversy---see \citep{greaves,lewis,lewis1,price}).

I conclude that the difference between one real world (and many
possible ones) on the one hand,  and many real worlds with
subjective uncertainty injected into them on the other hand, is
irrelevant for the justification of the form of Born rule. The
Born formula can be derived from the requirement that the measure
$\mu$ should be definable in terms of only $\psi$ (and should
therefore not depend on anything else, like information about
future events). The meaning of this $\mu$, be it in terms of modal
probabilities or subjective many-worlds uncertainty, is something
that cannot be derived but must be added in an interpretational
step.

Let us now direct our attention to a more specific comparison of
modal ideas and many worlds, relating to technical details.

One would perhaps expect the many worlds interpretation to follow
the earlier explained modal way of fixing the definite observables
in determining the worlds (each world corresponding to a different
set of values of the definite observables), because a basic idea
of the many-worlds interpretation is that the formalism is
self-sufficient and that knowledge of the state is enough to
obtain a full description of the universe. However, the dominant
opinion among many-world adherents is different, namely that a
privileged decomposition of the state is determined by the
dynamical mechanism of decoherence. At first sight this
decoherence recipe is almost identical to the one described in
sections 2 and 3. Indeed, decoherence leads to a state of the
general form
\begin{equation}
|\psi\rangle = \sum_{k} c_{k}|\psi_{k}\rangle\otimes
|E_{k}\rangle, \label{eq:decoherence} \end{equation} with
$|\psi_{k}\rangle$ representing the part of the world undergoing
decoherence, and $|E_{k}\rangle$ representing the decohering part
(usually the environment of the system that is undergoing
decoherence). This looks similar to Eq.\ \ref{eq:modal}. The
difference is that the states $|E_{k}\rangle$ are not exactly
orthogonal, so that Eq.\ \ref{eq:decoherence} is not a
biorthogonal decomposition. It is true that typically not only
$\langle E_{i}|E_{j}\rangle \rightarrow 0$ when $t\rightarrow
\infty$, but that $\langle E_{i}|E_{j}\rangle \simeq 0$, if $i\neq
j$, even very soon after the onset of the decoherence process:
this process is very effective. Still, this inner product can
never be assumed to really vanish at finite times. This means that
the branches corresponding to the terms in Eq.\
\ref{eq:decoherence} are not disjoint: there is interference
between them. In the case of a biorthogonal decomposition the
expectation value---taken in the total state---of any observable
of the form $A \otimes I$ is the sum of contributions from the
different branches; but this is not true in the case of the state
of Eq.\ \ref{eq:decoherence}. In the latter case there are
\textit{cross terms} in addition to the contributions from the
individual branches. Although these cross terms will typically be
tremendously small, they are important from a conceptual point of
view. They indicate that the total situation represented by the
state cannot be viewed as a juxtaposition of independent
alternatives or isolated worlds.

Another way of formulating the same point is that the Born
probabilities of measurement results of observables of the form $A
\otimes I$, calculated in the individual branches, when added with
weights equal to the Born probabilities of those individual
branches themselves, do not reproduce the probabilities of these
outcomes in the total state. This is a violation of a consistency
condition on the interpretation of $|c_k|^2$ as a probability both
within the individual worlds \emph{and} in the universe consisting
of many worlds.

A second conceptual difficulty is that the way the state
$|\psi\rangle$ has been decomposed in Eq.\ \ref{eq:decoherence} is
not unique. A change of basis from ${|\psi_{k}\rangle}$ to a
slightly different set of mutually orthogonal vectors will
preserve the general form of Eq.\ \ref{eq:decoherence}, together
with the almost orthogonality of the decohering states. So
decoherence does not lead to a well-defined set of branches.
Many-worlds proponents usually see this as an innocent form of
vagueness: it is sufficient that on the macroscopic level the
usual quantities, defined within observational precision, become
definite---this is compatible with some leeway in the quantities
that are definite on a microscopic scale. As \citet{butterfield}
formulates it:
\begin{quote}
``...the ubiquity and astonishing efficiency of decoherence means
that for all macrosystems ... the selected quantity will be very
nearly unique---so that the vagueness will be unnoticeable by the
standards of precision usual for macroscopic physics''.
 \end{quote}
This may seem plausible, but it is not a solid result backed up by
calculations. That is, at present there is no guarantee that the
different possible choices of ${|\psi_{k}\rangle}$ that make it
possible to write the state in the `decoherent form' of
(\ref{eq:decoherence}), with $\langle E_{i}|E_{j}\rangle$ very
small, are close to each other in Hilbert space. As \citet{baccia}
has shown in the context of modal interpretations, instabilities
may occur in the biorthogonal decomposition, and sometimes this
may lead to the selection of observables that are very different
from the usual macroscopic observables one would expect to be
selected. These results apply in particular to situations in which
there are very many degrees of freedom, for example when
macroscopic bodies are immersed in a decohering environment. It is
true that these results have been rigourously derived only for the
biorthogonal decomposition, but if $\langle E_{i}|E_{j}\rangle$ is
very small instead of exactly zero, one should expect a similar
behavior. This raises the general question of whether the
decoherence scheme will be capable of always defining an adequate
set of worlds, or definite outcomes of experiments. This is the
same question that can be asked in the case of modal
interpretations (cf.\ footnote 2).

In general, it seems that the technical difficulties which have
been suggested to exist for modal interpretations based on the
biorthogonal decomposition (in conditions with continuous or very
many degrees of freedom \citep{baccia,donald}), should also be
taken seriously in interpretations based on the idea that
decoherence singles out a preferred decomposition of the state.
These difficulties have been investigated in some detail only
within the modal framework, because here there are precise
mathematical rules that define the definite valued observables.
However, although there is more `slippage' in the decoherence
scheme, the problem of instability requires a solution in this
context as well. One should be careful here to distinguish between
two different questions. In standard treatments of decoherence one
starts by writing the state as a superposition of eigenstates of
an \textit{a priori} given preferred observable, often position;
and then shows that these eigenstates become correlated, through
the interaction with the surroundings, with almost orthogonal
environment states. The question we are facing here, however, is
whether the total state, after having undergone decoherence,
defines adequate physical quantities. The uncontroversial reply to
the first question does not answer the second; and it is to this
second question that the above questions pertain.

A final point is that the use of approximations, which is very
common in decoherence approaches, runs the risk of getting into
vicious circles. In particular, the motivation for neglecting
`small' components of $\psi$ seems to presuppose a probabilistic
interpretation of the coefficients with which these components
occur. But this would vitiate a \textit{deduction} of the Born
rule. Indeed, the Born rule or something equivalent to it would in
this case already be used in the definition of the definite events
to which the Born probabilities are later to be assigned.

In summary, both from the viewpoint of general considerations and
from that of more detailed quantum mechanical arguments there
seems little reason to prefer many-worlds interpretations over
modal ones. Decoherence is not an obvious help for the many-worlds
scheme---if the problems just mentioned can be solved, it still
remains unclear why decoherence would help the many-worlds
theorist and would not be available to adherents of modal ideas.

\section{Conclusion}

The Hilbert space formalism of quantum mechanics restricts
possible interpretations in the following sense: if we stipulate
that definite-valued quantities and a probability measure over
them should be definable from the quantum state and the Hilbert
space tensor product structure for the system and its environment,
this leads to unique expressions both for these definite
quantities and the measure. \textit{That} the state vector should
be thus interpreted in terms of definite quantities and
probabilities is not something that can be derived from the
mathematical formalism---it is an interpretational choice. Modal
interpretations implement this choice by postulating that the
quantum state represents possibilities of which only one is
realized in physical reality.

What the definite quantities turn out to be obviously depends on
what we stipulate about the elements in the formalism on which
these quantities are to depend. Here we have focused on what
follows from the requirement of definability from only the state
and the bipartite tensor product structure of Hilbert space. Other
stipulations are possible while staying within the general
framework of the modal programme, which is characterized by
objectivity (description of the world through objective physical
quantities) and a fundamental role for probability. These
alternative options deserve further investigation.

\end{document}